\documentclass[]{spie}  

\newlength\savewidth\newcommand\shline{\noalign{\global\savewidth\arrayrulewidth
\global\arrayrulewidth 1pt}\hline\noalign{\global\arrayrulewidth\savewidth}}
\usepackage{amsmath,amsfonts,amssymb}
\usepackage{graphicx}
\usepackage{multirow}

\usepackage[colorlinks=true, allcolors=blue]{hyperref}

\title{Swin transformers are robust to distribution and concept drift in endoscopy-based longitudinal rectal cancer assessment}

\author[a]{Jorge Tapias Gomez}
\author[a]{Aneesh Rangnekar}
\author[b]{Hannah Williams}
\author[b]{Hannah M. Thompson}
\author[b]{Julio Garcia-Aguilar}
\author[b]{Joshua Jesse Smith}
\author[a]{Harini Veeraraghavan}
\affil[a]{Medical Physics, Memorial Sloan Kettering Cancer Center, New York City, USA}
\affil[b]{Surgery, Colorectal Service, Memorial Sloan Kettering Cancer Center, New York City, USA}

\begin{document} 
\maketitle

\begin{abstract}
Endoscopic images are used at various stages of rectal cancer treatment starting from cancer screening and diagnosis, during treatment to assess response and toxicity from treatments such as colitis, and at follow-up to detect new tumor or local regrowth. However, subjective assessment is highly variable and can underestimate the degree of response in some patients, subjecting them to unnecessary surgery, or overestimating response that places patients at risk of disease spread. Advances in deep learning have shown the ability to produce consistent and objective response assessments for endoscopic images. However, methods for detecting cancers, regrowth, and monitoring response during the entire course of patient treatment and follow-up are lacking. This is because automated diagnosis and rectal cancer response assessment require methods that are robust to inherent imaging illumination variations and confounding conditions (blood, scope, blurring) present in endoscopy images as well as changes to the normal lumen and tumor during treatment. Hence, a hierarchical shifted window (Swin) transformer was trained to distinguish rectal cancer from normal lumen using endoscopy images. Swin, as well as two convolutional (ResNet-50, WideResNet-50), and the vision transformer architectures, were trained and evaluated on follow-up longitudinal images to detect LR on in-distribution (ID) private datasets as well as on out-of-distribution (OOD) public colonoscopy datasets to detect pre/non-cancerous polyps. Color shifts were applied using optimal transport to simulate distribution shifts. Swin and ResNet models were similarly accurate in the ID dataset. Swin was more accurate than other methods (follow-up: 0.84, OOD: 0.83), even when subject to color shifts (follow-up: 0.83, OOD: 0.87), indicating the capability to provide robust performance for longitudinal cancer assessment.   
\end{abstract}


\section{PURPOSE}
\label{sec:intro}  
Flexible sigmoidoscopy is a routinely used procedure to diagnose and monitor treatment response in cancers of the gastrointestinal (GI) tract, such as rectal cancers. However, accurate visual assessment of cancer treatment response from endoscopic images is highly variable due to large variations in cancer appearance and varying degrees of response during and after treatment\cite{felder2021}. Prior works have used automated video analysis to detect pre/non-cancerous lesions\cite{Mesejo2016,ali2022} and segment polyps\cite{Zhao2023}. However, videos are typically not acquired during routine treatment follow-up, thus reducing the available numbers of training data\cite{Williams2024}. Cancers also depict wide appearance changes on longitudinal imaging due to cancer response and recurrence as well as normal tissue changes from treatments\cite{thompson2023}. Finally, endoscopic images are also subject to wide illumination variations due to scope positioning, occlusions, specular reflections, intensity saturation, and presence of blood and stool, presenting challenges for robust analysis\cite{ALI2021102002}. As a result, it is important to develop classifiers that are robust to both in-distribution (ID; similar to training data) as well as out-of-distribution (OOD) datasets. 

Hence, the purpose and objective of this work was to develop a deep learning classifier to distinguish rectal cancer from no cancer from longitudinal endoscopy images acquired before, during, and following treatment. To assess the robustness of our models to imaging and disease conditions, our models were evaluated under two distinct conditions: (a) \textbf{covariate or distribution shift\/} \rm produced by modifying the color histogram of images using an optimal transport-based color normalization method\cite{ConvWassersteinDistances} and (b) \textbf{concept shift\/} \rm by evaluating the model trained with baseline, during, and restaging (within 8 weeks of completing treatment) on follow-up scans to detect local regrowth (LR) of tumors as well as entirely unrelated cohorts consisting of patients with non-cancerous conditions such as colitis, adenomas and polyps.  

In addition, our work also differs from prior works that predominantly focused on pre-cancerous polyp detection and classification\cite{yamada2019,turan2022,ALI2021102002,Dong2023} and segmentation\cite{fan2020pranet,Dong2023PolypPVT}. Our approach involves single-step analysis to classify images as containing tumors or normal lumen and does not require segmentation. We also used a hierarchical transformer (Swin) \cite{liu2021swin} as opposed to a vision transformer (ViT) previously used for segmentation tasks\cite{fan2020pranet,Dong2023PolypPVT}. 

An additional challenge for creating robust classifiers is the limited training examples, especially when videos are not available to access multiple video frames as training examples. Prior works addressed the same issue through weakly supervised training\cite{yamada2019}, synthesis of additional samples using generative adversarial networks\cite{turan2022}, as well as pretrained foundation models created with natural images and transferred to medical imaging tasks\cite{huix2024natural}. We employed the last approach by using pretrained models and combined training regularization schemes using data augmentation\cite{ramesh2023dissecting} and learning rate scheduling\cite{goyal2018accurate} to improve models' performance. 

Our key contributions and novelty include: (i) deep learning-based detection of rectal cancer longitudinally as the tumor evolves in appearance in response to treatment and at recurrence, (ii) a pretrained Swin transformer-based approach used for classification, and (iii) rigorous evaluation of the OOD robustness of Swin, ViT as well as the convolutional networks under distribution and concept drift using private and public datasets.

\section{METHODS}
\noindent\textbf{Data preprocessing: \/}\rm Standard-of-care white light endoscopy images were analyzed in all datasets. The images varied in size from 568 $\times$ 424 pixels to 1920 $\times$ 1080 pixels. For training and evaluation, we resized all images to 224 $\times$ 224 pixels. We normalized the images using the ImageNet dataset mean and standard deviation, [0.485, 0.456, 0.406] and [0.229, 0.224, 0.225], respectively\cite{Deng2009}. Institutional datasets were acquired using white-light flexible endoscopy with an Olympus scope (model CF-160S) from patients diagnosed with locally advanced rectal cancer (LARC) and who underwent treatment with total neoadjuvant therapy (TNT). 

\subsection{Datasets}
\noindent\textbf{In-distribution (ID) dataset \/}\rm consisted of 2,570 images from 200 different patients (1,329 with tumor and 1,241 without visible tumor) acquired at baseline or pre-TNT, interval or during-TNT, restaging appointments performed within 8 weeks of completing TNT (post-TNT), as well as during follow-up. All models were trained with an identical set of 1,337 images, validated on 448 images, and evaluated on a \underline{held out ID test set of 785 images}. In addition, 152 images of patients with LARC who showed a clinical complete response at restaging and were placed on watch and wait for surveillance and followed for up to 2 years (72 with complete response) or until they developed local regrowth (80 with LR) of the same cancer were evaluated after the model was "locked" for testing. 

\noindent\textbf{OOD dataset \/}\rm consisted of a subset of 50 images containing polyps (n = 31) and ulcerative colitis (n = 19) from the publicly available Hyper-Kvasir dataset\cite{Borgli2020}. The selected subset was chosen such that images contained confounding artifacts such as blood (90\% of cases), stool, blurring, and burned text annotations. This dataset was made available only after the model was `locked' for testing.

\subsection{Architecture details}
We trained four different architectures, consisting of two convolutional networks, namely, ResNet-50\cite{he2016deep} and Wide Residual Network (Wide ResNet-50)\cite{zagoruyko2016}, a hierarchical shifted window transformer (Swin)\cite{liu2021swin}, and a vision transformer (ViT)\cite{ViT}. ResNet-50 is comprised of 50 bottleneck residual blocks, WideResNet-50 had 50 layers with a widened architecture with a widening factor of 2, Swin used a Swin-Base with patch size 4 and window size 7 for an input resolution of 224 $\times$ 224 backbone while ViT used as ViT-Base model with 16 $\times$ 16 patch size as the backbone. Default architectures with pretrained model checkpoints produced by supervised training on the ImageNet\cite{Deng2009} dataset as provided by PyTorch were used to bootstrap network weights. The fully connected layers close to the output were replaced with a linear layer (2048 $\times$ 2), (2048 $\times$ 2), (1024 $\times$ 2), and (768 $\times$ 2) respectively. Models were fine-tuned to generate binary classification (tumor vs no-tumor) from the endoscopy images in the ID dataset.

\subsection{Color shift via Optimal Transport}
The color histogram of images was altered to assess the robustness of the models to changes in image distribution. Color shift was implemented using an optimal transport-based approach that treats images as three-dimensional distributions (1D luminance [L] and 2D histogram of chrominance [ab]) and computes illumination transformation without altering image content to match that of a reference image\cite{ConvWassersteinDistances}. The reference image was selected as the one devoid of tumor and free of artifacts (scope, light saturation) and confounders (blood, water bubbles, stool). 

\subsection{Implementation Details} 
All networks were trained for 50 epochs using a mini-batch of 8, with a learning rate of 2$e{-5}$, and the Adam optimizer with a linear weight decay of 1$e^{-4}$. Balanced sampling was used to reduce the class imbalance between tumor and no-tumor samples within the mini-batches. Learning rate scheduling and data augmentations, such as random image rotations in $90^\circ$ steps, random horizontal flips, and random vertical flips, all of which were found to be effective for colonoscopy video analysis were used to address data size limitations\cite{ramesh2023dissecting}. When using learning rate scheduling, we increased the learning rate from 0 to 2$e^{-5}$ within 10 epochs. We used Pytorch 1.13.1\cite{paszke2019pytorch} with NVIDIA GPUs for all our experiments.

\begin{figure} [h]
\begin{center}
\begin{tabular}{c} 
\includegraphics[height=3.5cm]{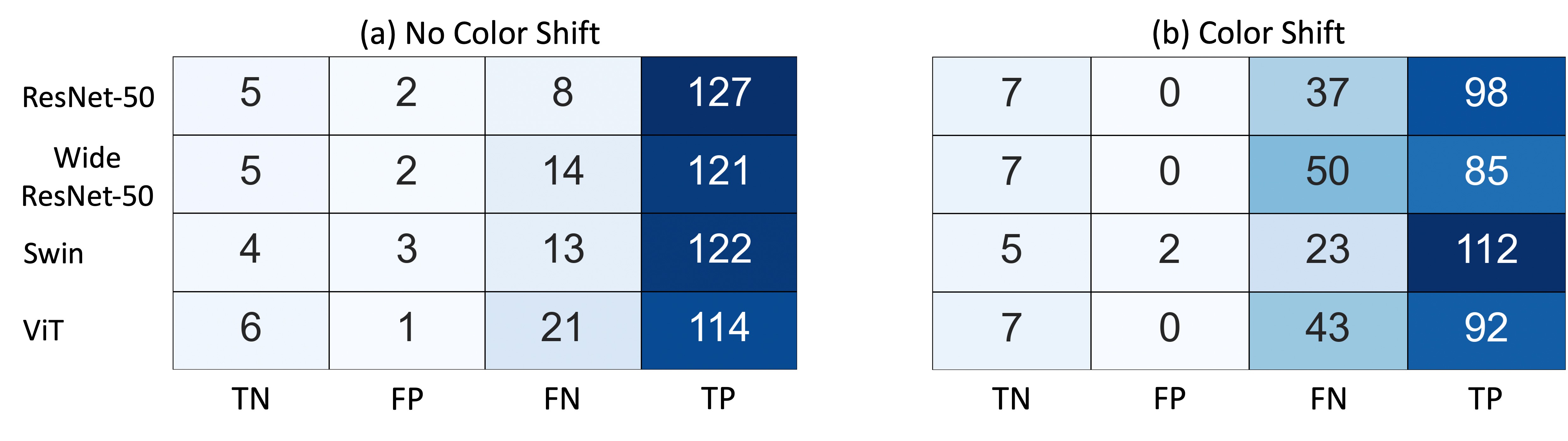}
\end{tabular}
\end{center}
\caption[example] 
{ \label{fig:tnt} 
Confusion matrix for classifying ID testing dataset by the analyzed DL models for images subjected to (a) no color shift and (b) with color shift. TN: true negatives, FP: false positives, FN: false negatives, and TP: true positives.}
\end{figure} 

\begin{table}[h]
\centering
\def\arraystretch{1.25}
\caption{Accuracy (Acc), sensitivity (Sens), and specificity (Spec) for classifying LR from no-tumor in follow-up images.}
\vspace{0.2cm}
\begin{tabular}{lccc}
\multirow{2}{*}{Model} & \multicolumn{3}{c}{No Color Shift} \\ \cline{2-4} 
 & Acc & Sens & Spec\\ \shline
ResNet-50 & 0.82 $\pm$ 0.01 & 0.61 $\pm$ 0.02  & 0.95 $\pm$ 0.01  \\
Wide ResNet-50  & 0.81 $\pm$ 0.02 & 0.53 $\pm$ 0.07 & 0.97 $\pm$ 0.01 \\
Swin & 0.84 $\pm$ 0.02  & 0.63 $\pm$ 0.07  & 0.96 $\pm$ 0.01  \\
ViT & 0.81 $\pm$ 0.04  & 0.63 $\pm$ 0.05  & 0.95 $\pm$ 0.02  \\ \hline
\end{tabular}%
\hspace{0.5cm}
\begin{tabular}{ccc}
\multicolumn{3}{c}{Color Shift} \\ \hline
Acc & Sens & Spec\\ \shline
0.83 $\pm$ 0.02 & 0.56 $\pm$ 0.04  & 0.98 $\pm$ 0.01 \\
0.81 $\pm$ 0.02 & 0.47 $\pm$ 0.08 & 0.99 $\pm$ 0.01\\
0.83 $\pm$ 0.02 & 0.60 $\pm$ 0.06  & 0.95 $\pm$ 0.01  \\
0.79 $\pm$ 0.03  & 0.53 $\pm$ 0.04  & 0.92 $\pm$ 0.03  \\ \hline
\end{tabular}%
\label{tab:post+lr}
\end{table}

\begin{table}[h]
\centering
\def\arraystretch{1.25}

\caption{Accuracy (Acc), Sensitivity (Sens), and Specificity (Spec) for classification in the external dataset (concept drift).}
\vspace{0.2cm}
\begin{tabular}{lccc}
\multirow{2}{*}{Model} & \multicolumn{3}{c}{No Color Shift} \\ \cline{2-4} 
 & Acc & Sens & Spec\\ \shline
ResNet-50 & 0.74 $\pm$ 0.03 & 0.95 $\pm$ 0.03  & 0.3 $\pm$ 0.11 \\
Wide ResNet-50  & 0.77 $\pm$ 0.04 & 0.93 $\pm$ 0.01 & 0.42 $\pm$ 0.11\\
Swin & 0.83 $\pm$ 0.02 & 0.96 $\pm$ 0.04  & 0.54 $\pm$ 0.13 \\
ViT & 0.74 $\pm$ 0.04 & 1.00 $\pm$ 0.00  & 0.17 $\pm$ 0.11 \\ \hline
\end{tabular}%
\hspace{0.1cm}
\begin{tabular}{ccc}
\multicolumn{3}{c}{Color Shift} \\ \hline
Acc & Sens & Spec\\ \shline
0.78 $\pm$ 0.01 & 0.87 $\pm$ 0.02  & 0.58 $\pm$ 0.00  \\
0.72 $\pm$ 0.06 & 0.72 $\pm$ 0.16 & 0.70 $\pm$ 0.18\\
0.87 $\pm$ 0.02 & 0.96 $\pm$ 0.05  & 0.68 $\pm$ 0.16  \\
0.83 $\pm$ 0.01 & 0.98 $\pm$ 0.03  & 0.51 $\pm$ 0.03 \\ \hline
\end{tabular}%
\label{tab:external}
\end{table}

\section{Results}
\label{sec:results}

\noindent\textbf{Performance on ID dataset \/}\rm comprising of pre-TNT and during-TNT images is shown as a confusion matrix in Fig. \ref{fig:tnt}. All models showed similar performance when evaluated on images with no color shift, with ResNet-50 producing the best true positive and least number of false negatives, followed by Swin. On the other hand, Swin was the most accurate when evaluated on the same dataset subjected to color shift. These results indicate that Swin is more resistant to color shifts in the ID dataset compared to the ResNet-50 model.

Table~\ref{tab:post+lr} shows the performance of the same four models evaluated on the follow-up images. True detections corresponded to LR that differ in appearance from the tumors occurring in baseline or residual tumors at restaging. Appearance at restaging often is confounded by radiation and chemotherapy-induced changes to normal tissues. As shown, all models were similarly accurate especially when evaluated on data without color shifts. Performance differences emerged when analyzed with color-shifted images, where Swin produced the best sensitivity followed by ResNet-50. ViT was the least accurate on all metrics in the color-shifted cases. 

\noindent\textbf{Robustness to concept drift: \/}\rm Table~\ref{tab:external} shows the performance of the various models with the public dataset. As shown, Swin was the most accurate method in both analysis scenarios and showed a clear performance improvement with an accuracy margin of 0.06 to 0.09 compared to other methods and the lowest standard deviation of 0.02 when evaluated on color-shifted images. As opposed to the ID-dataset (Table.~\ref{tab:post+lr}) where both ResNet-50 and Swin were similar, the latter model showed a clear accuracy gain over the former model.

\begin{figure} [hb]
\begin{center}
\begin{tabular}{c} 
\includegraphics[height=7cm]{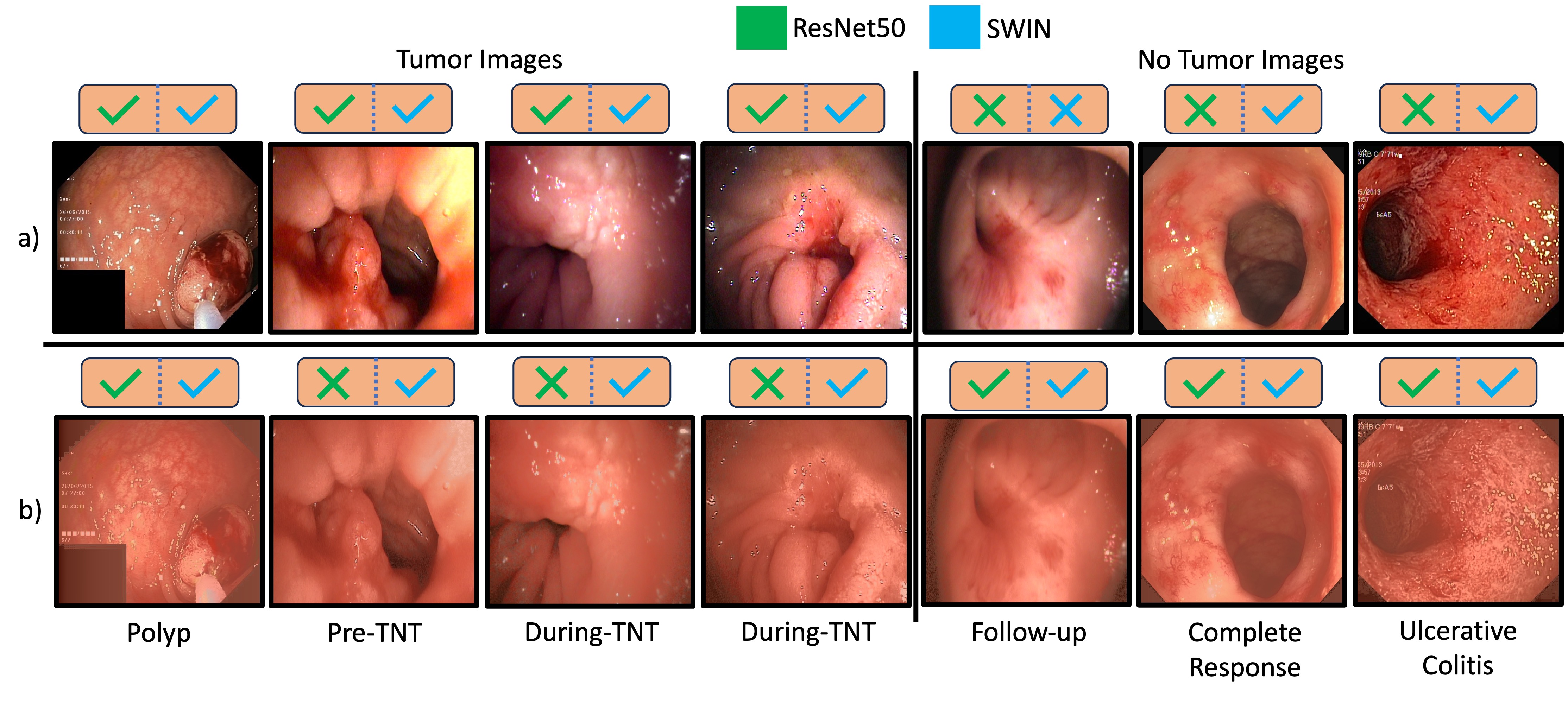}
\end{tabular}
\end{center}
\caption[example] 
{ \label{fig:example} 
Effect of color shifting on ResNet-50 and Swin predictions in both tumor and no-tumor images at different stages of treatment and the external dataset.}
\end{figure}

\noindent\textbf{Analysis of color and concept shift: \/}\rm Fig. \ref{fig:example} shows multiple representative example images from the OOD as well as ID dataset taken at various time points before, during, restaging, and follow-up without (Fig. \ref{fig:example} (a)) and with (Fig. \ref{fig:example}(b)) color shift when evaluated with ResNet50 and the Swin models. As shown, the ResNet50 model can correctly identify polyps and tumors at various stages of treatment when images were not subjected to color shifts but shows the variable ability to correctly identify tumors when the images were subjected to color shifts. However, ResNet-50 was positively impacted in correctly identifying no-tumor (including on colitis cases) when images were color-shifted. On the other hand, Swin is robust to image differences produced by color shift for both tumor and no-tumor cases, indicating robust performance under both color and concept shifts.  
\section{Conclusions}

We trained and evaluated multiple deep learning classifiers to distinguish tumor from no-tumor using endoscopy images taken at various stages of treatment and follow-up in patients with LARC. Our analysis showed that Swin was robust to both color and concept drifts compared to convolutional network-based ResNet-50 as well as the ViT approach. Validation on larger multi-institutional cohorts and a more in-depth analysis of the confounders such as blood, stool, specularities, and treatment effects on performance is planned for future work.

\section{Compliance with Ethical Standards}
The available images were collected retrospectively and correlated with clinical information. This study was approved by the institutional review board of Memorial Sloan Kettering Cancer Center.

\section{Acknowledgements}

This research was supported by the Department of Surgery at Memorial Sloan Kettering. We thank Maria Widmar, Iris H Wei, Emmanouil P Pappou, Garrett M Nash, Martin R Weiser, and Philip B Paty for their assistance in collecting endoscopic images.

\bibliography{report} 

\begin{thebibliography}{10}

\bibitem{felder2021}
Felder, S., Patil, S., Kennedy, E., and Garcia-Aguilar, J., ``Endoscopic feature and response reproducibility in tumor assessment after neoadjuvant therapy for rectal adenocarcinoma,'' {\em Ann Surg Oncol}~{\bf 28}(9),  5205--5223 (2021).

\bibitem{Mesejo2016}
Mesejo, P., Pizarro, D., Abergel, A., Rouquette, O., Beorchia, S., Poincloux, L., and Bartoli, A., ``Computer-aided classification of gastrointestinal lesions in regular colonoscopy,'' {\em IEEE Transactions on Medical Imaging}~{\bf 35}(9),  2051--2063 (2016).

\bibitem{ali2022}
Ali, S., ``Where do we stand in ai for endoscopic image analysis? deciphering gaps and future directions,'' {\em NPJ Digital Medicine}~{\bf 5}(184) (2022).

\bibitem{Zhao2023}
Wang, Z., Liu, C., Zhang, S., and Dou, Q., ``Foundation model for endoscopy video analysis via large-scale self-supervised pre-train,'' in [{\em Medical Image Computing and Computer Assisted Intervention -- MICCAI 2023}{\nolinebreak\hspace{0.1em}]},   101--111, Springer Nature Switzerland, Cham (2023).

\bibitem{Williams2024}
Williams, H., Thompson, H., Lee, C., Rangnekar, A., Gomez, J., Widmar, M., Wei, I., Pappou, E., Nash, G., Weiser, M., Paty, P., Smith, J., Veeraraghavan, H., and Garcia-Aguilar, J., ``Assessing endoscopic response in locally advanced rectal cancer treated with total neoadjuvant therapy: Development and validation of a highly accurate convolutional neural network,'' {\em Ann Surg Oncol}  (2024).

\bibitem{thompson2023}
Thompson, H., Kim, J., Jimenez-Rodriguez, R., Garcia-Aguilar, J., and Veeraraghavan, H., ``Deep learning-based model for identifying tumors in endoscopic images from patients with locally advanced rectal cancer treated with total neoadjuvant therapy,'' {\em Dis Colon Rectum}~{\bf 66}(3),  383--391 (2023).

\bibitem{ALI2021102002}
Ali, S., Dmitrieva, M., Ghatwary, N., Bano, S., Polat, G., Temizel, A., Krenzer, A., Hekalo, A., Guo, Y.~B., Matuszewski, B., Gridach, M., Voiculescu, I., Yoganand, V., Chavan, A., Raj, A., Nguyen, N.~T., Tran, D.~Q., Huynh, L.~D., Boutry, N., Rezvy, S., Chen, H., Choi, Y.~H., Subramanian, A., Balasubramanian, V., Gao, X.~W., Hu, H., Liao, Y., Stoyanov, D., Daul, C., Realdon, S., Cannizzaro, R., Lamarque, D., Tran-Nguyen, T., Bailey, A., Braden, B., East, J.~E., and Rittscher, J., ``Deep learning for detection and segmentation of artefact and disease instances in gastrointestinal endoscopy,'' {\em Medical Image Analysis}~{\bf 70},  102002 (2021).

\bibitem{ConvWassersteinDistances}
Solomon, J., de~Goes, F., Peyr\'{e}, G., Cuturi, M., Butscher, A., Nguyen, A., Du, T., and Guibas, L., ``Convolutional wasserstein distances: Efficient optimal transportation on geometric domains,'' {\em ACM Trans. Graph.}~{\bf 34}(4) (2015).

\bibitem{yamada2019}
Yamada, M., Saito, Y., Imaoka, H., Saiko, M., Yamada, S., Kondo, H., Takamaru, H., Sakamoto, T., Sese, J., Kuchiba, A., Shibata, T., and Hamamoto, R., ``Development of a real-time endoscopic image diagnosis support system using deep learning technology in colonoscopy,'' {\em Sci Rep} (14465) (2019).

\bibitem{turan2022}
Turan, M. and Durmus, F., ``\textsc{UC}-\textsc{N}f\textsc{N}et: Deep learning-enabled assessment of ulcerative colitis from colonoscopy images,'' {\em Med Image Anal}~{\bf 82},  102587 (2022).

\bibitem{Dong2023}
Dong, Z., Wang, J., Li, Y., Deng, Y., Zhou, W., Zeng, X., Gong, D., Liu, J., Pan, J., Shang, R., Xu, Y., Xu, M., Zhang, L., Zhang, M., Tao, X., Zhu, Y., Du, H., Lu, Z., Yao, L., Wu, L., and Yu, H., ``Explainable artificial intelligence incorporated with domain knowledge diagnosing early gastric neoplasms under white light endoscopy,'' {\em NPJ Digit Med}~{\bf 6}(1),  64 (2023).

\bibitem{fan2020pranet}
Fan, D.-P., Ji, G.-P., Zhou, T., Chen, G., Fu, H., Shen, J., and Shao, L., ``Pranet: Parallel reverse attention network for polyp segmentation,'' in [{\em International conference on medical image computing and computer-assisted intervention}{\nolinebreak\hspace{0.1em}]},   263--273, Springer (2020).

\bibitem{Dong2023PolypPVT}
Dong, B., Wang, W., Fan, D.-P., Li, J., Fu, H., and Shao, L., ``Polyp-pvt: Polyp segmentation with pyramid vision transformers,'' {\em CAAI Artificial Intelligence Research}~{\bf 2},  9150015 (2023).

\bibitem{liu2021swin}
Liu, Z., Lin, Y., Cao, Y., Hu, H., Wei, Y., Zhang, Z., Lin, S., and Guo, B., ``Swin transformer: Hierarchical vision transformer using shifted windows,'' in [{\em Proceedings of the IEEE/CVF international conference on computer vision}{\nolinebreak\hspace{0.1em}]},   10012--10022 (2021).

\bibitem{huix2024natural}
Huix, J.~P., Ganeshan, A.~R., Haslum, J.~F., S{\"o}derberg, M., Matsoukas, C., and Smith, K., ``Are natural domain foundation models useful for medical image classification?,'' in [{\em Proceedings of the IEEE/CVF Winter Conference on Applications of Computer Vision}{\nolinebreak\hspace{0.1em}]},   7634--7643 (2024).

\bibitem{ramesh2023dissecting}
Ramesh, S., Srivastav, V., Alapatt, D., Yu, T., Murali, A., Sestini, L., Nwoye, C.~I., Hamoud, I., Sharma, S., Fleurentin, A., Exarchakis, G., Karargyris, A., and Padoy, N., ``Dissecting self-supervised learning methods for surgical computer vision,'' (2023).

\bibitem{goyal2018accurate}
Goyal, P., Dollár, P., Girshick, R., Noordhuis, P., Wesolowski, L., Kyrola, A., Tulloch, A., Jia, Y., and He, K., ``Accurate, large minibatch sgd: Training imagenet in 1 hour,'' (2018).

\bibitem{Deng2009}
Deng, J., Dong, W., Socher, R., Li, L.-J., Li, K., and Fei-Fei, L., ``Imagenet: A large-scale hierarchical image database,'' in [{\em 2009 IEEE Conference on Computer Vision and Pattern Recognition}{\nolinebreak\hspace{0.1em}]},   248--255 (2009).

\bibitem{Borgli2020}
Borgli, H., Thambawita, V., Smedsrud, P.~H., Hicks, S., Jha, D., Eskeland, S.~L., Randel, K.~R., Pogorelov, K., Lux, M., Nguyen, D. T.~D., Johansen, D., Griwodz, C., Stensland, H.~K., Garcia-Ceja, E., Schmidt, P.~T., Hammer, H.~L., Riegler, M.~A., Halvorsen, P., and de~Lange, T., ``{HyperKvasir, a comprehensive multi-class image and video dataset for gastrointestinal endoscopy},'' {\em Scientific Data}~{\bf 7}(1),  283 (2020).

\bibitem{he2016deep}
He, K., Zhang, X., Ren, S., and Sun, J., ``Deep residual learning for image recognition,'' in [{\em Proceedings of the IEEE conference on computer vision and pattern recognition}{\nolinebreak\hspace{0.1em}]},   770--778 (2016).

\bibitem{zagoruyko2016}
Zagoruyko, S. and Komodakis, N., ``Wide residual networks,'' in [{\em Proceedings of the British Machine Vision Conference 2016, {BMVC} 2016, York, UK, September 19-22, 2016}{\nolinebreak\hspace{0.1em}]},  {BMVA} Press (2016).

\bibitem{ViT}
Dosovitskiy, A., Beyer, L., Kolesnikov, A., Weissenborn, D., Zhai, X., Unterthiner, T., Dehghani, M., Minderer, M., Heigold, G., Gelly, S., Uszkoreit, J., and Houlsby, N., ``An image is worth 16x16 words: Transformers for image recognition at scale,'' in [{\em 9th International Conference on Learning Representations, {ICLR} 2021, Virtual Event, Austria, May 3-7, 2021}{\nolinebreak\hspace{0.1em}]},  OpenReview.net (2021).

\bibitem{paszke2019pytorch}
Paszke, A., Gross, S., Massa, F., Lerer, A., Bradbury, J., Chanan, G., Killeen, T., Lin, Z., Gimelshein, N., Antiga, L., et~al., ``Pytorch: An imperative style, high-performance deep learning library,'' {\em Advances in neural information processing systems}~{\bf 32} (2019).

\end{thebibliography}
\bibliographystyle{spiebib} 

\end{document}